\documentclass[%
 aip,
 rsi,
 amsmath,amssymb,
]{revtex4-2}

\usepackage{graphicx}
\usepackage{dcolumn}
\usepackage{bm}
\usepackage[utf8]{inputenc}
\usepackage[T1]{fontenc}
\usepackage{mathptmx}
\usepackage{graphicx}
\newcommand{\fig}[1]{Fig.\,\ref{#1}}
\newcommand{\Fig}[1]{Figure\,\ref{#1}}
\newcommand{\eq}[1]{Eq.\,(\ref{#1})}
\newcommand{\resxx}{\makebox{Re$(\sigma_{xx})$}}

\usepackage{color,soul}

\usepackage{cleveref}
\begin{document}

\preprint{AIP/123-QED}

\title[Resistively loaded coplanar waveguide]{Resistively loaded coplanar waveguide for microwave measurements of induced carriers }

\author{M. L. Freeman}
\email{MatthewLyleFreeman@outlook.com}
 \altaffiliation[Also at ]{Physics Department, Florida Sate University}
 \affiliation{ 
 	National High Magnetic Field Laboratory, Tallahassee, Florida 32310, USA
 }%
\author{Tzu-Ming Lu}%
\affiliation{Sandia National Laboratories, Albuquerque, New Mexico 87185, USA}
\affiliation{Center for Integrated Nanotechnologies, Sandia National Laboratories, Albuquerque, New Mexico 87123, USA}

\author{L. W. Engel}
\email{engel@magnet.fsu.edu}
 \affiliation{ 
 	National High Magnetic Field Laboratory, Tallahassee, Florida 32310, USA
}%

\date{\today}

\begin{abstract}
 	We describe the use of a coplanar waveguide whose slots are filled with resistive film, a resistively loaded coplanar waveguide (RLCPW), to measure two dimensional electron systems (2DES).  The RLCPW applied to the sample hosting the 2DES provides a uniform metallic surface serving as a gate, to control the areal charge density of the 2DES.  As a demonstration of this technique we present measurements on a Si MOSFET, and a model that successfully converts microwave transmission coefficients into conductivity of a nearby 2DES capacitively coupled to the RLCPW.  We also describe the process of fabricating the highly resistive metal film required for fabrication of the RLCPW.

\end{abstract}

\maketitle

\section{Introduction }
	Coplanar  waveguide (CPW) is a standard and well-characterized type of microwave transmission line, which has the advantage of being constructed on a single plane.  A CPW, as shown in \fig{mosschm}a and b, is constructed of metal film forming a driven center conductor and broad ground planes, separated by slots. Placing a CPW in capacitive or direct contact with a two-dimensional electron system (2DES) allows broadband measurements of absorption and phase shift due to the 2DES.  CPW-based microwave spectroscopy of this sort has been performed on 2DES hosted in semiconductors \cite{leimw,murthyrvw,hatkebi,stone,drichko} and on large-area graphene \cite{dragoman,skulason,wuadv}.  In these measurements the transmission through the CPW is particularly sensitive to the conductivity of the 2DES under the slot regions of the CPW. 

	In studying 2DES it  is often necessary  to vary the density of the system by means of biasing a metallic gate.  Some semiconductor-hosted 2DES, as well as high quality graphene and transition metal dichalcogenides are not charged by deliberately placed dopants, but rely on charge induced by a gate.  Such gates, if in close proximity to the 2DES, create complications for CPW based microwave measurements, because a highly conducting gate screens the microwave electric field, reducing the sensitivity to changes in the conductivity of the 2DES.  One approach to this problem  has been to fabricate gates spaced away from the CPW and the 2DES, and biasing these gates with high voltage.  This method  \cite{hatkebi} was performed with a front gate, spaced by a vacuum (or air)  gap and a thin glass plate above the CPW in \fig{mosschm}a, and also with gates on the back side of a semiconductor wafer (far below the CPW in the figure).  When the spaced front gate is used, the density in the slots and under the CPW metal is mismatched unless the 2DES is biased precisely  relative to the CPW to match the the density in the slots; such matching is difficult in practice. Both the back and spaced front gates have limited range of density tunability, require high bias voltage, and do not work in some samples.  Another approach, which uses a highly conductive gate combined with a sensitive density modulation technique to detect cyclotron resonance, is described by ref. \onlinecite{stone}. 

	In this paper we demonstrate  microwave measurements using a resistively loaded CPW (RLCPW), as shown in \fig{mosschm}c and d, in which a   resistive film gate is placed  under the highly conductive CPW conductors, and only partly screens the microwave field.  This is analogous to the use of semitransparent gates in optical experiments.  Our motivation for this is experimentation in heterojunction insulated gate field effect transistors (HIGFETS)\cite{kanehig,willetthig,sarkozyhig,panhig,panhigberry} which are semiconductor devices known to give particularly low disorder and wide density tunability.   
	However, for the proof-of-concept demonstration in this paper we present data on a Si  metal-oxide-semiconductor field-effect transistor (MOSFET), a well-understood device whose large disorder is expected to render its conductivity nearly independent of frequency in our measuring range.

	The RLCPW retains the advantages of wide bandwidth and coplanar (single-surface) construction, but presents a uniform surface to the sample.  In semiconductors this removes density gradiant due to Schottky potentials, and reduces strain due to differential contraction relative to a conventional CPW.  Microwave measurements can couple capacitively to samples, so in that case  are essentially contactless, though a single contact to a sample (possibly only at elevated temperature) is required to induce the carriers. The contact, which can be challenging in undoped devices, can be  highly resistive, with a leak of $10^{12} \ \Omega$ or more sufficient to populate most samples.
	A limitation of the RLCPW measurement is that it  becomes insensitive if the 2DES conductivity is much less than that of the resistive film, so fabricating a high resistivity film in the RLCPW is an important aspect of the design.  In the following we describe fabrication and measurement of an RLCPW with high resistivity metal film on the Si MOSFET.  In addition we present an analytical model for conversion of microwave transmission measurement to a 2DES conductivity for the capacitively coupled case, and compare its results to the measurements. 

\section{Experiment set-up and MOSFET device}
\begin{figure}
	 	\includegraphics[width=0.9\linewidth]{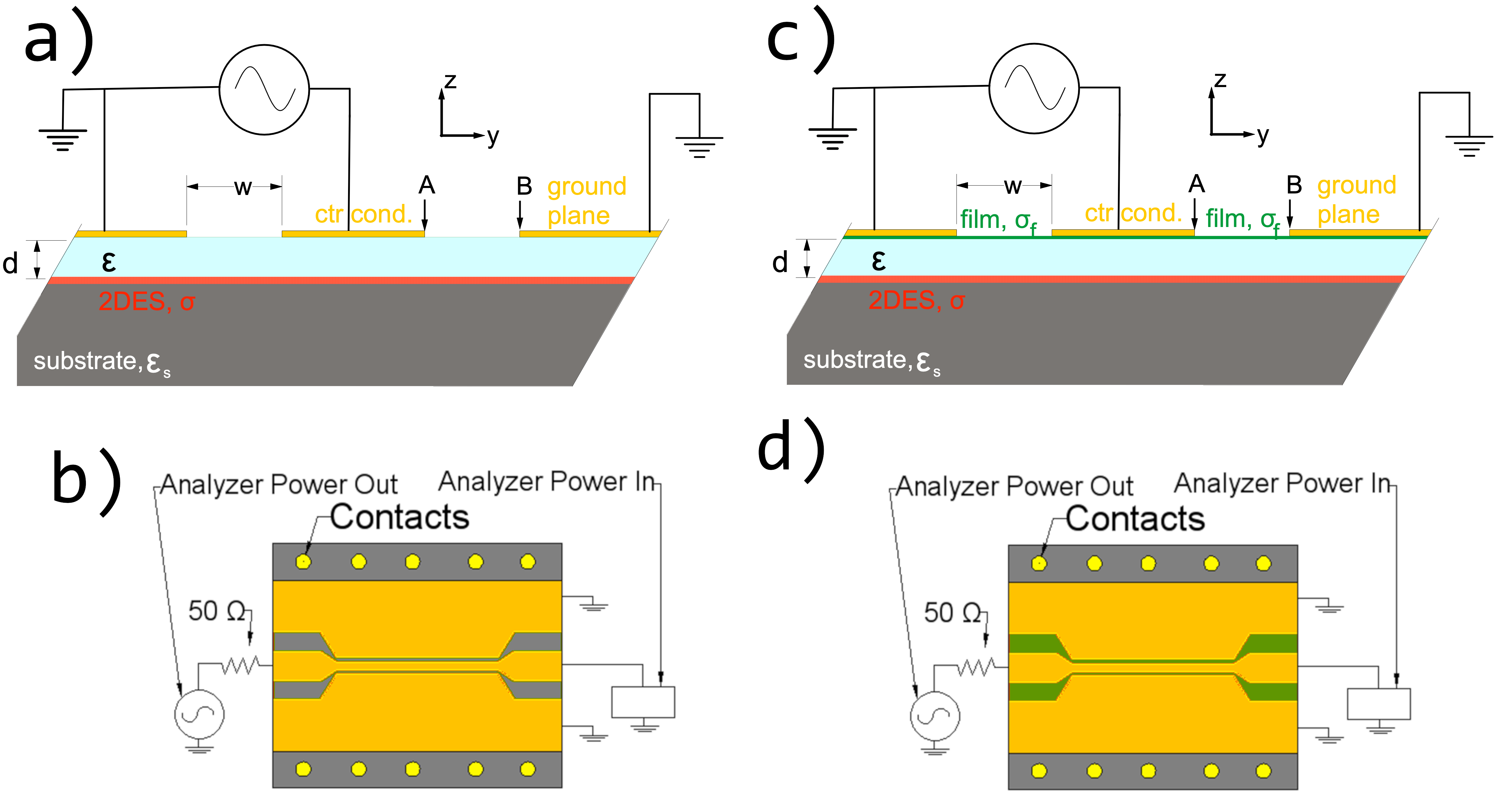}  
		\caption{Coplanar waveguide (CPW) and resistively loaded coplanar waveguide (RLCPW), on top of a two-dimensional electron system (2DES) hosted in a semiconductor.  The 2DES is separated from the CPW or RLCPW by a dielectric layer shown as light blue which is $d$ thick and has dielectric constant $\epsilon$.  The highly conductive center conductor and ground planes are shown as gold and the 2DES as red.  The slot width is $w$.  In the RLCPW case of (c) the resistive film that bridges the RLCPW slots is shown as green.  (a) shows a CPW in cross section.  (b) shows a schematic of CPW in top view.  (c) and (d) show RLCPW schematics in cross section and top view respectively.  The model in the text refers to points A and B, which mark the edges of the center conductor and side plane at the slot.  \label{mosschm}}
	\end{figure}

	\Cref{mosschm}d shows a schematic of the experiment with a top view of the RLCPW, which we fabricated as a gate on a MOSFET with oxide thickness of 35 nm.  The edges of the  RLCPW are tapered to facilitate a low-reflectance connection to microstrip lines on kapton film, which are in turn connected to coaxial cables.  The length of the RLCPW between the tapers is 2.5 mm, the center conductor width is 45 $\mu$m, and the slot width ($w$) is 30 $\mu$m.  The sample is located in a 0.3 K cryostat, and is connected to a room-temperature network analyzer, which serves as transmitter and receiver.  Ohmic contacts on the device enable dc measurements of the MOSFET, and allow the 2D channel to be populated by the gate.    
   
	For the purpose of charging the channel, the RLCPW can be regarded as a uniform gate. (The microwave voltage applied between the RLCPW center line and its ground planes is much smaller than the dc bias voltages.)  For convenience of the microwave connections, the RLCPW  is  grounded to the cryostat, and we apply negative bias ($-V_b$) to the contacts of the MOSFET relative to that ground to populate the channel.  The device substrate is lightly p-type doped with a room temperature resistivity of 1-10 $\Omega\cdot cm$.  This lightly doped substrate freezes out at low temperatures and minimizes microwave absorption at our measuring temperature of 0.3 K.  The properties of the oxide layer of the MOSFET are important for the model calculation of conductivity from microwave absorption: the MOSFET had oxide thickness of 35 nm and we took the oxide dielectric constant to be 3.7.
  
	\subsection* {Resistive film}
		A key component of the RLCPW is the resistive film. 
		Here we used a recently developed thin film composed of evaporated Ge, Ni, and Pt layers\cite{Harris2021}.  The thicknesses of the Ge, Ni, and Pt layers were 20, 0.5, and 1 nm, respectively.  After deposition, the film was annealed in a forming gas environment at 525 C for 10 sec.  This thermal treatment created a resistive film with a resistivity on the order of 1-10 k$\Omega$/square \cite{Harris2021}.  In this work, we  obtained  the 2D resistivity of the film, $\rho_f \approx 3390 \ \Omega / $square, from the dc resistance between the center conductor and the ground planes, making use of the much higher conductivity of the side planes and center conductor.   
 
\section{Experimental Results}
\begin{figure}
		\includegraphics[width=5in]{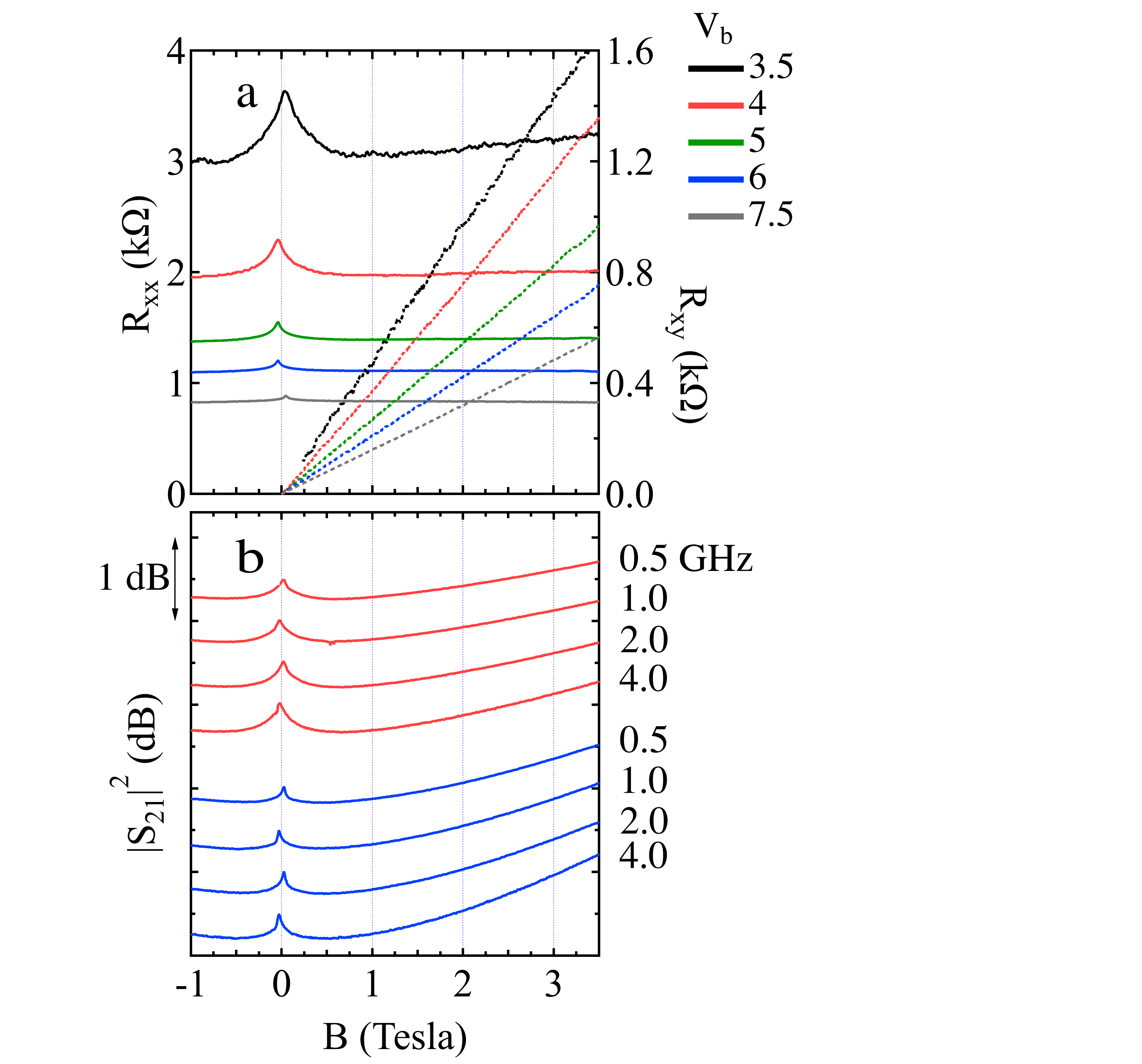}
		\caption{Field dependence at various gate bias voltages, $V_b$. (a) DC resistances vs magnetic field.  Solid lines are longitudinal resistance $R_{xx}(k\Omega)$ and dashed lines are Hall resistance $R_{xy}(k\Omega)$.  (b) Transmission coefficient $|s_{21}|^2$ (dB) vs magnetic field at various frequencies. Upper (red) group of curves is for $V_b=4$ V, lower (blue) is for $V_b=6$ V. The curves are offset vertically for clarity.  \label{bswp}}
	\end{figure}

	Using the ohmic contacts, we characterize the MOSFET by standard techniques of dc transport, for different $V_b$ as magnetic field ($B$) is varied.  \Fig{bswp}a shows the diagonal and Hall resistances, $R_{xx}$ and $R_{xy}$ vs $B$, for several bias voltages. As $V_b$ is increased, $R_{xx}$ decreases and the slope of the Hall resistance decreases due to increasing 2D density, $n$ (shown below in Fig. 4a).    The  Hall resistance follows the classical formula $R_{xy}=B/ne$.  Prominent in these and subsequent  traces vs magnetic field is a peak at $B=0$ due to weak localization; the  apparent slight displacement  of this peak  from $B=0$ in  some traces is an artifact of remanent field in the superconducting magnet. 

	\Fig{bswp}b shows the microwave transmission $|s_{21}|^2$ at various frequencies ($f$), for $V_b=4$ and $6$ V.  Unlike the dc $R_{xx}$ curves of panel a, the curves exhibit an upward curvature which arises because the transmission is sensitive to $\sigma_{xx}$ rather than $R_{xx}$. The $|s_{21}|^2$ traces are offset from each other for clarity, and their $B$ dependence changes little for the different frequencies.  Measurements with long thin transmission lines are nearly insensitive to $\sigma_{xy}$, as has been shown for measurements in the quantum Hall regime \cite{leimw}. 

\section{Analytical Model  }
	\begin{figure}
	 	\includegraphics[width=0.4\linewidth]{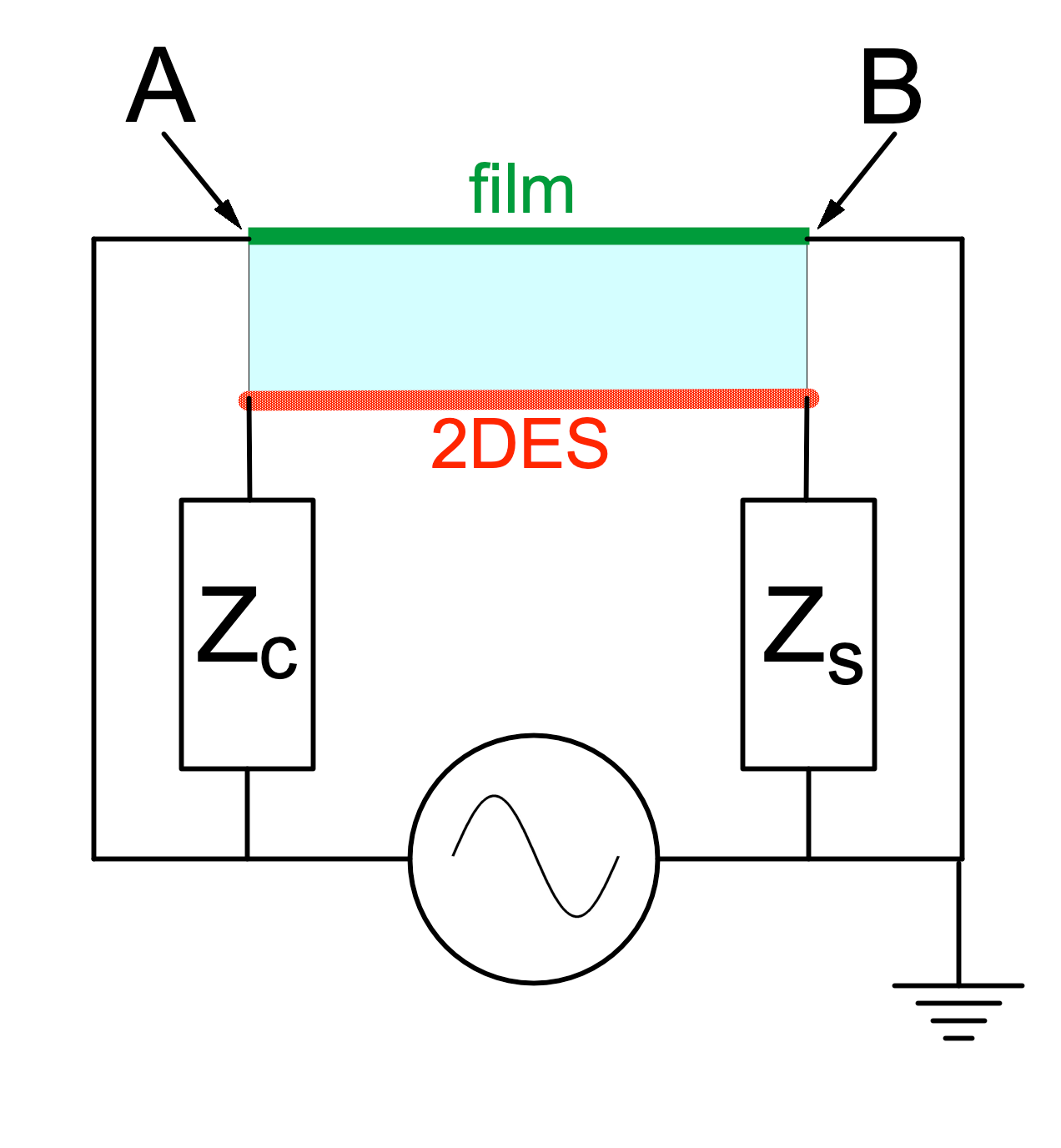}
		\caption{Circuit model, with $Z_c$ the impedance from the center conductor coupled to the 2DES, and $Z_s$ the impedance from the side plane to the 2DES.  A and B are the same points as in Fig. 1.  $Z_c$ and $Z_s$ are per unit length in the propagation direction, and the 2DES and film are coupled capacitively.  \label{circuit}}
	\end{figure}

	The model presented here can be used to  calculate complex transmission coefficient $s_{21}$ \cite{adam}  from a 2DES conductivity, given the dimensions of the RLCPW, the film resistivity and the per unit area capacitance between the 2DES and the film.  To convert measured $s_{21}$ to 2DES conductivity, the model must be inverted, which is conveniently done numerically.  A more time-consuming alternative to the model is to use commercially available simulator software, such as Sonnet \cite{sonnet}, which we used to check the analytical model.   

 	\Fig{mosschm}b shows a schematic cross section of the RLCPW system narrowly spaced above a 2DES.  We model the system in the quasi-TEM approximation  which is applicable to situations where the wavelength is much larger than the RLCPW slot width.  Besides high frequency effects which become important  when the wavelength approaches $w$, the model neglects edge effects\cite{foglerhuse} and also the  capacitance due to lines of force which terminate at both ends on the 2DES; both these effects are small for the RLCPW with achievable film resistivities.  For $w>>d$, even with the smallest achievable film conductivity $\sigma_f$, these effects are not important.  Quantum (density of states) capacitance\cite{eisenstein} is neglected in this  paper and is not important for the MOSFET device, but could be incorporated in the coupling capacitance when appropriate.    
 
	We consider a distributed circuit, composed of parallel plate transmission lines comprised of the 2DES and the metal above; this approach has been used elsewhere\cite{mehrotra,burke}.  In the present case the parallel plate lines are transverse ($y$ direction in \fig{mosschm}) to the RLCPW propagation direction ($x$ in the figure, pointing toward the viewer of the figure) of the RLCPW.  The parallel plate lines, whose top layer is the RLCPW and whose bottom layer is the 2DES have voltage and current related by
	\begin{equation} 
 		V_1'=-Z_1 I_1,\ \   V_2'=-\sigma^{-1} I_1,\ \ I_1'=(V_2-V_1)i\omega C_c,\ \ I_2'=(V_1-V_2)i\omega C_c,  \label{diff}
 	\end{equation}
 	where primes indicate differentiation in $y$.  $V_1$, $I_1$ pertain to the top layer RLCPW, and $V_2$, $I_2$ to the 2DES.  $C_c$ is the capacitance per unit area of the dielectric layer, given by $C_c=\epsilon/d$, where $d$ is the  dielectric thickness.  $Z_1$ is taken to be zero for the highly conductive center conductor and side planes, and $\sigma_f^{-1}$ for the region of the slots.    
 
	We calculate the complex transmission coefficient $s_{21}$ \cite{adam} using the circuit of \fig{circuit} to obtain the load admittance $Y_L$ per unit length in the propagation direction, presented to the RLCPW by the resistive film and by the capacitively coupled 2DES.  
	In the figure, the impedances $Z_s=( 1/i\sigma\omega C_c)^{1/2}$, $Z_c=Z_s \coth [(i\omega C_c/\sigma)^{1/2} a/2]$, result from \eq{diff} applied to a side plane and the center conductor regions.  The coupled resistive film and 2DES is between the points marked A and B in the figure.  Here $Z_s$ is taken for a wide side plane, assuming current vanishes within the ground plane as $y$ gets farther from the outer edge of the slot, and would need to be modified to include the outside boundary of the side plane in the case \cite{mehrotra, burke} of low loss and large kinetic inductance. 
  

	\begin{equation}
		\begin{bmatrix}
			1\\
			0\\
			0\\
			0
		\end{bmatrix}=
		\begin{bmatrix}
			0 & 1/2& \alpha/2 &  \alpha /2 \\
			Z_c\sigma  & 0 & Z_c +2\beta & Z_c -2\beta \\
			e^{-\gamma w}  w &e^{-\gamma w}  &-e^{-2\gamma w}\alpha&   \alpha \\
			e^{-\gamma w}  Z_s\sigma  &0 & e^{-2\gamma w}(Z_s-2\beta) & Z_s +2\beta
		\end{bmatrix}
		\begin{bmatrix}
			c_1 \\
			c_2 \\
			c_3\\
			c_4
		\end{bmatrix},
	\end{equation}
	With $\bar{Z}=(\sigma_f^{-1}+\sigma^{-1})/2$, $\gamma=(2i\omega C_c \bar{Z})^{1/2}$, $\beta= \bar{Z}/\gamma$, and $\alpha=(\gamma\sigma_f)^{-1} $.  Of the coefficients $c_k$, obtained from solving this system, only $c_1$ is of interest, and gives the admittance due to the sample layer, per unit length in the propagation direction of the RLCPW, as $Y_L= -c_1(\sigma_f+\sigma)$.  

	We design the RLCPW in the quasi-TEM approximation such that if the sample layer and the resistive film  were absent, the underlying CPW with characteristic impedance $Z_0=50 \ \Omega$ and phase velocity $v_{p0}$ would result \cite{ghionenaldi}.  (Presently, this task can easily be performed for a substrate of known dielectric constant using online calculators \cite{onlinecpw}.)  The model for the RLCPW incorporates $Y_L$ into the admittance per unit length in the propagation direction resulting in propagation constant
	$\Gamma_R=[i\omega   (i\omega +Y_L Z_0 v_{p0}  )  ]^{1/2}/v_{p0} $ and chararacteristic impedance
	$z_R=(1-iY_L Z_0v_{p0}/\omega)^{-1/2}$. $z_R$ is normalized for a 50 $\Omega$ measuring system, in which the complex transmission coefficient of the RLCPW is 
	\begin{equation}
		s_{21}=\frac{ z_R }{   (z_R^2+1)\cosh(\Gamma_Rl) +2z_R \sinh(\Gamma_R l)  }.
	\end{equation}
 
 	\begin{figure}
 		\includegraphics[width=3in]{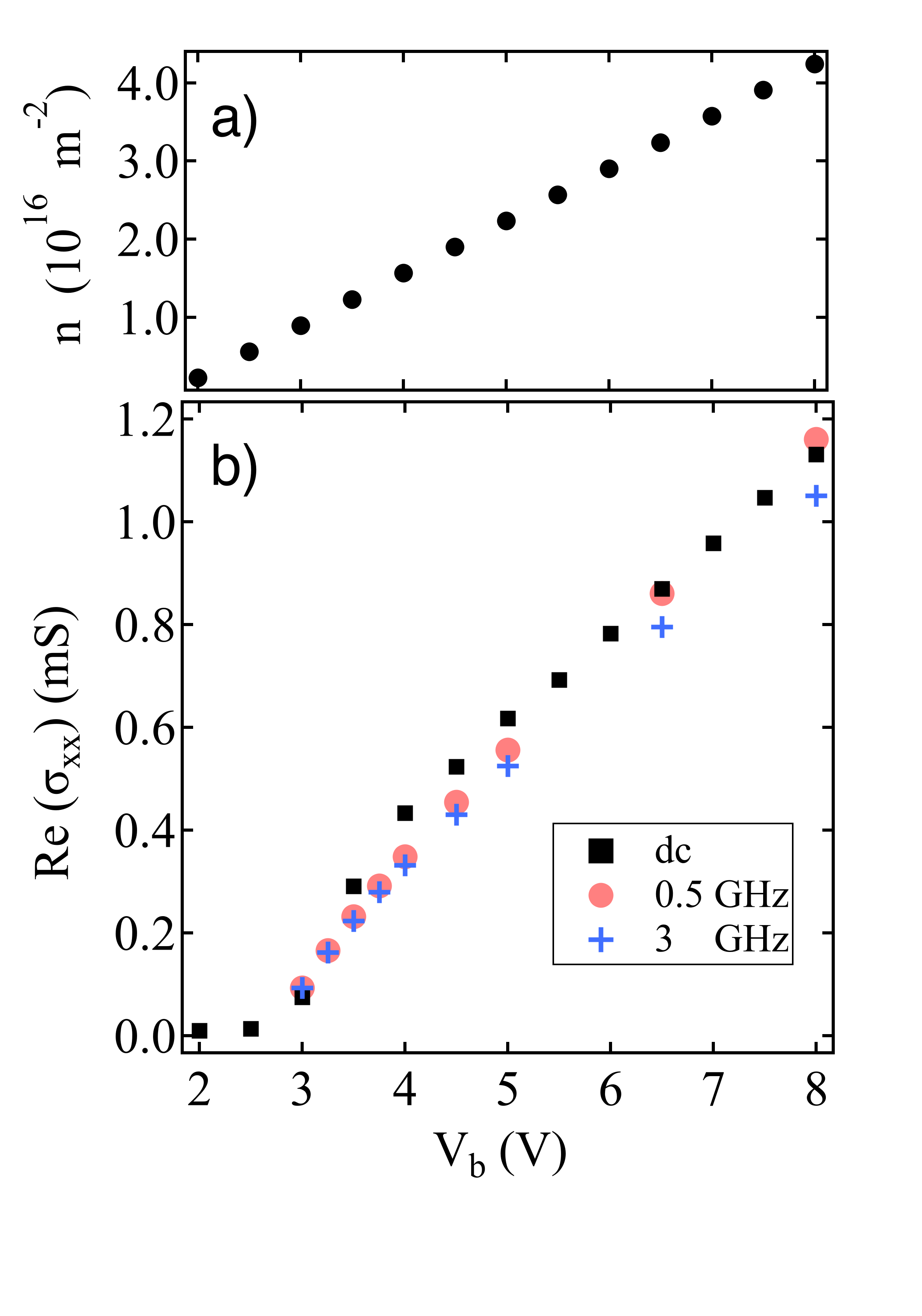}
 		\caption{(a) MOSFET areal density, $n$, vs bias voltage, $V_b$. (b) Real diagonal conductivity of the MOSFET, Re$(\sigma_{xx})$ vs $V_b$ measured at dc and calculated by means of the model (see text) from   normalized $|s_{21}|^2$ at 0.5 and 3 GHz.  \label{compare}}
	\end{figure}

\section{Microwave conductivity measurement with the RLCPW}
	We demonstrate microwave conductivity measurement with the RLCPW by comparing microwave and dc conductivities, relying on the independence of the conductivity of the MOSFET on the frequency.  The disorder in the MOSFET leads to a short carrier momentum relaxation time, $\tau$. Frequency independence is expected when $ 2\pi f \tau\ll 1$. Using $\tau\sim\sigma_{xx}m^*/ne^2$, where $\sigma_{xx}$ is taken at zero frequency and field, and for Si the relevant effective mass $m^*\sim 0.19$ free electron masses\cite{sze}, we find $\tau\lesssim 2.2 \times 10^{-13}$ seconds.  Hence $\omega\tau\ll1$ is satisfied for our measuring conditions ($f\le 4$ GHz).   

	We use the dc data of \fig{bswp}a to test obtaining  the 2DES conductivity $\sigma_{xx}$ from microwave measurements.  The dc resistivity, $\rho_{xx}$ is calculated by the van der Pauw method \cite{seeger} from resistances obtained from different sets of ohmic contacts at the edges of the RLCPW.  We find the dc $\sigma_{xx}$ from the tensor inversion, $\sigma_{xx}=\rho_{xx}/(\rho_{xx}^2 +\rho_{xy}^2)$, where the part of $R_{xy}$ that is antisymmetric about $B=0$ is $\rho_{xy}$.

	\Fig{compare}b shows real diagonal conductivities \resxx, vs $V_b$,  obtained from  dc measurements through the contacts, and  from $s_{21}$ measurements at frequencies of 0.5 and 3 GHz. To avoid the sharp $B=0$ weak localization feature in the conductivity,  the data were taken with a 2 T magnetic field.   To remove the frequency dependent effects  of cables,  the   0.5 and 3 GHz   \resxx\  data shown were obtained with a simplified calibration procedure, assuming zero absorption from the 2DES at zero bias.   We measured $s_{21}(V_b)/s_{21}(V_b=0)$, and iteratively compared  it to the results of the model of the previous section for  varying $\sigma$, $s_{21}(\sigma)/s_{21}(\sigma=0)$, taking the imaginary part of $\sigma$ to be zero, as is appropriate for $ 2\pi f \tau\ll 1$.   The dc and microwave \resxx\ data are grouped together, demonstrating our main point that the RLCPW method can be used to obtain  $\sigma_{xx}$ with appropriate modeling.  There are several sources of discrepancies between the points of the three traces on the graph.  First,   dc resistivity measurements used to calculate \resxx\ probe different areas on the MOSFET  than the RLCPW  measurement, which  reports  conductivity as an average mainly from the areas under the RLCPW slots.  The differences between the two microwave frequency  data at larger \resxx\ may be due to  the increase in reflection from the RLCPW which is not removed by the simple calibration procedure.

	In summary, we have demonstrated that microwave-frequency   measurements can be carried out with an RLCPW transmission line as a sample is gated with a resistive film to vary induced carrier density widely.  We have shown that  the conductivity of the gated 2DES can be obtained quantitatively from such measurements, by means of an analytic model into which the resistivity of the film loading the RLCPW is incorporated. 
 
	\newpage

\begin{acknowledgments}
	The microwave measurements and sample design at NHMFL were supported by Department of Energy (Grant No. DE- FG02-05-ER46212).
	The National High Magnetic Field Laboratory is supported by the National Science Foundation Cooperative Agreement No. DMR-1644779 and the State of Florida.
	This work was partly funded by the U.S.~Department of Energy (DOE) Office of Basic Energy Sciences (BESs). 
	This work was performed, in part, at the Center for Integrated Nanotechnologies, a U.S.~DOE, Office of BESs, user facility.  
	Sandia National Laboratories is a multimission laboratory managed and operated by National Technology and Engineering Solutions of Sandia LLC, a wholly owned subsidiary of Honeywell International Inc.~for the U.S.~DOE's National Nuclear Security Administration under contract DE-NA0003525. 
	This paper describes objective technical results and analysis. 
	Any subjective views or opinions that might be expressed in the paper do not necessarily represent the views of the U.S. DOE or the United States Government.  
	The data that support the findings of this study are available from the corresponding authors upon reasonable request. 
\end{acknowledgments}

\bibliography{mattmosB-TL}

\begin{thebibliography}{24}%
\makeatletter
\providecommand \@ifxundefined [1]{%
 \@ifx{#1\undefined}
}%
\providecommand \@ifnum [1]{%
 \ifnum #1\expandafter \@firstoftwo
 \else \expandafter \@secondoftwo
 \fi
}%
\providecommand \@ifx [1]{%
 \ifx #1\expandafter \@firstoftwo
 \else \expandafter \@secondoftwo
 \fi
}%
\providecommand \natexlab [1]{#1}%
\providecommand \enquote  [1]{``#1''}%
\providecommand \bibnamefont  [1]{#1}%
\providecommand \bibfnamefont [1]{#1}%
\providecommand \citenamefont [1]{#1}%
\providecommand \href@noop [0]{\@secondoftwo}%
\providecommand \href [0]{\begingroup \@sanitize@url \@href}%
\providecommand \@href[1]{\@@startlink{#1}\@@href}%
\providecommand \@@href[1]{\endgroup#1\@@endlink}%
\providecommand \@sanitize@url [0]{\catcode `\\12\catcode `\$12\catcode
  `\&12\catcode `\#12\catcode `\^12\catcode `\_12\catcode `\%12\relax}%
\providecommand \@@startlink[1]{}%
\providecommand \@@endlink[0]{}%
\providecommand \url  [0]{\begingroup\@sanitize@url \@url }%
\providecommand \@url [1]{\endgroup\@href {#1}{\urlprefix }}%
\providecommand \urlprefix  [0]{URL }%
\providecommand \Eprint [0]{\href }%
\providecommand \doibase [0]{https://doi.org/}%
\providecommand \selectlanguage [0]{\@gobble}%
\providecommand \bibinfo  [0]{\@secondoftwo}%
\providecommand \bibfield  [0]{\@secondoftwo}%
\providecommand \translation [1]{[#1]}%
\providecommand \BibitemOpen [0]{}%
\providecommand \bibitemStop [0]{}%
\providecommand \bibitemNoStop [0]{.\EOS\space}%
\providecommand \EOS [0]{\spacefactor3000\relax}%
\providecommand \BibitemShut  [1]{\csname bibitem#1\endcsname}%
\let\auto@bib@innerbib\@empty
\bibitem [{\citenamefont {Engel}\ \emph {et~al.}(1993)\citenamefont {Engel},
  \citenamefont {Shahar}, \citenamefont {Kurdak},\ and\ \citenamefont
  {Tsui}}]{leimw}%
  \BibitemOpen
  \bibfield  {author} {\bibinfo {author} {\bibfnamefont {L.~W.}\ \bibnamefont
  {Engel}}, \bibinfo {author} {\bibfnamefont {D.}~\bibnamefont {Shahar}},
  \bibinfo {author} {\bibfnamefont {{\c C}.}~\bibnamefont {Kurdak}},\ and\
  \bibinfo {author} {\bibfnamefont {D.~C.}\ \bibnamefont {Tsui}},\ }\bibfield
  {title} {\enquote {\bibinfo {title} {Microwave frequency dependence of
  integer quantum \mbox{{H}all} effect: Evidence for finite-frequency
  scaling},}\ }\href {https://doi.org/10.1103/PhysRevLett.71.2638} {\bibfield
  {journal} {\bibinfo  {journal} {Phys. Rev. Lett.}\ }\textbf {\bibinfo
  {volume} {71}},\ \bibinfo {pages} {2638--2641} (\bibinfo {year}
  {1993})}\BibitemShut {NoStop}%
\bibitem [{\citenamefont {Sambandamurthy}\ \emph {et~al.}(2006)\citenamefont
  {Sambandamurthy}, \citenamefont {Wang}, \citenamefont {Lewis}, \citenamefont
  {Chen}, \citenamefont {Engel}, \citenamefont {Tsui}, \citenamefont
  {Pfeiffer},\ and\ \citenamefont {West}}]{murthyrvw}%
  \BibitemOpen
  \bibfield  {author} {\bibinfo {author} {\bibfnamefont {G.}~\bibnamefont
  {Sambandamurthy}}, \bibinfo {author} {\bibfnamefont {Z.}~\bibnamefont
  {Wang}}, \bibinfo {author} {\bibfnamefont {R.~M.}\ \bibnamefont {Lewis}},
  \bibinfo {author} {\bibfnamefont {Y.~P.}\ \bibnamefont {Chen}}, \bibinfo
  {author} {\bibfnamefont {L.~W.}\ \bibnamefont {Engel}}, \bibinfo {author}
  {\bibfnamefont {D.~C.}\ \bibnamefont {Tsui}}, \bibinfo {author}
  {\bibfnamefont {L.~N.}\ \bibnamefont {Pfeiffer}},\ and\ \bibinfo {author}
  {\bibfnamefont {K.~W.}\ \bibnamefont {West}},\ }\bibfield  {title} {\enquote
  {\bibinfo {title} {Pinning mode resonances of new phases of 2d electron
  systems in high magnetic fields},}\ }\href@noop {} {\bibfield  {journal}
  {\bibinfo  {journal} {Solid State Commun.}\ }\textbf {\bibinfo {volume}
  {140}},\ \bibinfo {pages} {100 -- 106} (\bibinfo {year} {2006})}\BibitemShut
  {NoStop}%
\bibitem [{\citenamefont {Hatke}\ \emph {et~al.}(2015)\citenamefont {Hatke},
  \citenamefont {Liu}, \citenamefont {Engel}, \citenamefont {Shayegan},
  \citenamefont {Pfeiffer}, \citenamefont {West},\ and\ \citenamefont
  {Baldwin}}]{hatkebi}%
  \BibitemOpen
  \bibfield  {author} {\bibinfo {author} {\bibfnamefont {A.~T.}\ \bibnamefont
  {Hatke}}, \bibinfo {author} {\bibfnamefont {Y.}~\bibnamefont {Liu}}, \bibinfo
  {author} {\bibfnamefont {L.~W.}\ \bibnamefont {Engel}}, \bibinfo {author}
  {\bibfnamefont {M.}~\bibnamefont {Shayegan}}, \bibinfo {author}
  {\bibfnamefont {L.~N.}\ \bibnamefont {Pfeiffer}}, \bibinfo {author}
  {\bibfnamefont {K.~W.}\ \bibnamefont {West}},\ and\ \bibinfo {author}
  {\bibfnamefont {K.~W.}\ \bibnamefont {Baldwin}},\ }\bibfield  {title}
  {\enquote {\bibinfo {title} {Microwave spectroscopic studies of the bilayer
  electron solid states at low {L}andau filling in a wide quantum well},}\
  }\href {https://doi.org/10.1038/ncomms5154} {\bibfield  {journal} {\bibinfo
  {journal} {Nat. Commun.}\ }\textbf {\bibinfo {volume} {6}},\ \bibinfo {pages}
  {7071} (\bibinfo {year} {2015})}\BibitemShut {NoStop}%
\bibitem [{\citenamefont {Stone}\ \emph {et~al.}(2012)\citenamefont {Stone},
  \citenamefont {Du}, \citenamefont {Manfra}, \citenamefont {Pfeiffer},\ and\
  \citenamefont {West}}]{stone}%
  \BibitemOpen
  \bibfield  {author} {\bibinfo {author} {\bibfnamefont {K.}~\bibnamefont
  {Stone}}, \bibinfo {author} {\bibfnamefont {R.~R.}\ \bibnamefont {Du}},
  \bibinfo {author} {\bibfnamefont {M.~J.}\ \bibnamefont {Manfra}}, \bibinfo
  {author} {\bibfnamefont {L.~N.}\ \bibnamefont {Pfeiffer}},\ and\ \bibinfo
  {author} {\bibfnamefont {K.~W.}\ \bibnamefont {West}},\ }\bibfield  {title}
  {\enquote {\bibinfo {title} {Millimeter wave transmission spectroscopy of
  gated two-dimensional hole systems},}\ }\href
  {https://doi.org/10.1063/1.4711772} {\bibfield  {journal} {\bibinfo
  {journal} {Applied Physics Letters}\ }\textbf {\bibinfo {volume} {100}},\
  \bibinfo {pages} {192104} (\bibinfo {year} {2012})}\BibitemShut {NoStop}%
\bibitem [{\citenamefont {Drichko}\ \emph {et~al.}(2014)\citenamefont
  {Drichko}, \citenamefont {Diakonov}, \citenamefont {Malysh}, \citenamefont
  {Smirnov}, \citenamefont {Galperin}, \citenamefont {Ilyinskaya},
  \citenamefont {Usikova}, \citenamefont {Kummer},\ and\ \citenamefont {von
  K{\"a}nel}}]{drichko}%
  \BibitemOpen
  \bibfield  {author} {\bibinfo {author} {\bibfnamefont {I.~L.}\ \bibnamefont
  {Drichko}}, \bibinfo {author} {\bibfnamefont {A.~M.}\ \bibnamefont
  {Diakonov}}, \bibinfo {author} {\bibfnamefont {V.~A.}\ \bibnamefont
  {Malysh}}, \bibinfo {author} {\bibfnamefont {I.~Y.}\ \bibnamefont {Smirnov}},
  \bibinfo {author} {\bibfnamefont {Y.~M.}\ \bibnamefont {Galperin}}, \bibinfo
  {author} {\bibfnamefont {N.~D.}\ \bibnamefont {Ilyinskaya}}, \bibinfo
  {author} {\bibfnamefont {A.~A.}\ \bibnamefont {Usikova}}, \bibinfo {author}
  {\bibfnamefont {M.}~\bibnamefont {Kummer}},\ and\ \bibinfo {author}
  {\bibfnamefont {H.}~\bibnamefont {von K{\"a}nel}},\ }\bibfield  {title}
  {\enquote {\bibinfo {title} {Contactless measurement of alternating current
  conductance in quantum hall structures},}\ }\href
  {https://doi.org/10.1063/1.4898737} {\bibfield  {journal} {\bibinfo
  {journal} {Journal of Applied Physics}\ }\textbf {\bibinfo {volume} {116}},\
  \bibinfo {pages} {154309} (\bibinfo {year} {2014})}\BibitemShut {NoStop}%
\bibitem [{\citenamefont {Dragoman}\ \emph {et~al.}(2011)\citenamefont
  {Dragoman}, \citenamefont {Neculoiu}, \citenamefont {Cismaru}, \citenamefont
  {Muller}, \citenamefont {Deligeorgis}, \citenamefont {Konstantinidis},
  \citenamefont {Dragoman},\ and\ \citenamefont {Plana}}]{dragoman}%
  \BibitemOpen
  \bibfield  {author} {\bibinfo {author} {\bibfnamefont {M.}~\bibnamefont
  {Dragoman}}, \bibinfo {author} {\bibfnamefont {D.}~\bibnamefont {Neculoiu}},
  \bibinfo {author} {\bibfnamefont {A.}~\bibnamefont {Cismaru}}, \bibinfo
  {author} {\bibfnamefont {A.~A.}\ \bibnamefont {Muller}}, \bibinfo {author}
  {\bibfnamefont {G.}~\bibnamefont {Deligeorgis}}, \bibinfo {author}
  {\bibfnamefont {G.}~\bibnamefont {Konstantinidis}}, \bibinfo {author}
  {\bibfnamefont {D.}~\bibnamefont {Dragoman}},\ and\ \bibinfo {author}
  {\bibfnamefont {R.}~\bibnamefont {Plana}},\ }\bibfield  {title} {\enquote
  {\bibinfo {title} {Coplanar waveguide on graphene in the range 40 mhz--110
  ghz},}\ }\href {https://doi.org/10.1063/1.3615289} {\bibfield  {journal}
  {\bibinfo  {journal} {Applied Physics Letters}\ }\textbf {\bibinfo {volume}
  {99}},\ \bibinfo {pages} {033112} (\bibinfo {year} {2011})}\BibitemShut
  {NoStop}%
\bibitem [{\citenamefont {Skulason}\ \emph {et~al.}(2012)\citenamefont
  {Skulason}, \citenamefont {Nguyen}, \citenamefont {Guermoune}, \citenamefont
  {Siaj}, \citenamefont {Caloz},\ and\ \citenamefont {Szkopek}}]{skulason}%
  \BibitemOpen
  \bibfield  {author} {\bibinfo {author} {\bibfnamefont {H.~S.}\ \bibnamefont
  {Skulason}}, \bibinfo {author} {\bibfnamefont {H.~V.}\ \bibnamefont
  {Nguyen}}, \bibinfo {author} {\bibfnamefont {A.}~\bibnamefont {Guermoune}},
  \bibinfo {author} {\bibfnamefont {M.}~\bibnamefont {Siaj}}, \bibinfo {author}
  {\bibfnamefont {C.}~\bibnamefont {Caloz}},\ and\ \bibinfo {author}
  {\bibfnamefont {T.}~\bibnamefont {Szkopek}},\ }\bibfield  {title} {\enquote
  {\bibinfo {title} {Contactless impedance measurement of large-area
  high-quality graphene},}\ }in\ \href
  {https://doi.org/10.1109/MWSYM.2012.6259711} {\emph {\bibinfo {booktitle}
  {2012 IEEE/MTT-S International Microwave Symposium Digest}}}\ (\bibinfo
  {year} {2012})\ pp.\ \bibinfo {pages} {1--3}\BibitemShut {NoStop}%
\bibitem [{\citenamefont {Wu}\ \emph {et~al.}(2016)\citenamefont {Wu},
  \citenamefont {Wu}, \citenamefont {Kang}, \citenamefont {Chen}, \citenamefont
  {Li}, \citenamefont {Chen},\ and\ \citenamefont {Xu}}]{wuadv}%
  \BibitemOpen
  \bibfield  {author} {\bibinfo {author} {\bibfnamefont {Y.}~\bibnamefont
  {Wu}}, \bibinfo {author} {\bibfnamefont {Y.}~\bibnamefont {Wu}}, \bibinfo
  {author} {\bibfnamefont {K.}~\bibnamefont {Kang}}, \bibinfo {author}
  {\bibfnamefont {Y.}~\bibnamefont {Chen}}, \bibinfo {author} {\bibfnamefont
  {Y.}~\bibnamefont {Li}}, \bibinfo {author} {\bibfnamefont {T.}~\bibnamefont
  {Chen}},\ and\ \bibinfo {author} {\bibfnamefont {Y.}~\bibnamefont {Xu}},\
  }\bibfield  {title} {\enquote {\bibinfo {title} {Characterization of cvd
  graphene permittivity and conductivity in micro-/millimeter wave frequency
  range},}\ }\href {https://doi.org/10.1063/1.4963140} {\bibfield  {journal}
  {\bibinfo  {journal} {AIP Advances}\ }\textbf {\bibinfo {volume} {6}},\
  \bibinfo {pages} {095014} (\bibinfo {year} {2016})}\BibitemShut {NoStop}%
\bibitem [{\citenamefont {Kane}, \citenamefont {Pfeiffer},\ and\ \citenamefont
  {West}(1995)}]{kanehig}%
  \BibitemOpen
  \bibfield  {author} {\bibinfo {author} {\bibfnamefont {B.~E.}\ \bibnamefont
  {Kane}}, \bibinfo {author} {\bibfnamefont {L.~N.}\ \bibnamefont {Pfeiffer}},\
  and\ \bibinfo {author} {\bibfnamefont {K.~W.}\ \bibnamefont {West}},\
  }\bibfield  {title} {\enquote {\bibinfo {title} {High mobility \mbox{GaAs}
  heterostructure field effect transistor for nanofabrication in which dopant
  induced disorder is eliminated},}\ }\href {https://doi.org/10.1063/1.114391}
  {\bibfield  {journal} {\bibinfo  {journal} {Applied Physics Letters}\
  }\textbf {\bibinfo {volume} {67}},\ \bibinfo {pages} {1262--1264} (\bibinfo
  {year} {1995})}\BibitemShut {NoStop}%
\bibitem [{\citenamefont {Willett}\ \emph {et~al.}(2007)\citenamefont
  {Willett}, \citenamefont {Manfra}, \citenamefont {Pfeiffer},\ and\
  \citenamefont {West}}]{willetthig}%
  \BibitemOpen
  \bibfield  {author} {\bibinfo {author} {\bibfnamefont {R.~L.}\ \bibnamefont
  {Willett}}, \bibinfo {author} {\bibfnamefont {M.~J.}\ \bibnamefont {Manfra}},
  \bibinfo {author} {\bibfnamefont {L.~N.}\ \bibnamefont {Pfeiffer}},\ and\
  \bibinfo {author} {\bibfnamefont {K.~W.}\ \bibnamefont {West}},\ }\bibfield
  {title} {\enquote {\bibinfo {title} {Mesoscopic structures and
  two-dimensional hole systems in fully field effect controlled
  heterostructures},}\ }\href {https://doi.org/10.1063/1.2757128} {\bibfield
  {journal} {\bibinfo  {journal} {Applied Physics Letters}\ }\textbf {\bibinfo
  {volume} {91}},\ \bibinfo {eid} {033510} (\bibinfo {year}
  {2007})}\BibitemShut {NoStop}%
\bibitem [{\citenamefont {Sarkozy}\ \emph {et~al.}(2009)\citenamefont
  {Sarkozy}, \citenamefont {Gupta}, \citenamefont {Siegert}, \citenamefont
  {Ghosh}, \citenamefont {Pepper}, \citenamefont {Farrer}, \citenamefont
  {Beere}, \citenamefont {Ritchie},\ and\ \citenamefont {Jones}}]{sarkozyhig}%
  \BibitemOpen
  \bibfield  {author} {\bibinfo {author} {\bibfnamefont {S.}~\bibnamefont
  {Sarkozy}}, \bibinfo {author} {\bibfnamefont {K.~D.}\ \bibnamefont {Gupta}},
  \bibinfo {author} {\bibfnamefont {C.}~\bibnamefont {Siegert}}, \bibinfo
  {author} {\bibfnamefont {A.}~\bibnamefont {Ghosh}}, \bibinfo {author}
  {\bibfnamefont {M.}~\bibnamefont {Pepper}}, \bibinfo {author} {\bibfnamefont
  {I.}~\bibnamefont {Farrer}}, \bibinfo {author} {\bibfnamefont {H.~E.}\
  \bibnamefont {Beere}}, \bibinfo {author} {\bibfnamefont {D.~A.}\ \bibnamefont
  {Ritchie}},\ and\ \bibinfo {author} {\bibfnamefont {G.~A.~C.}\ \bibnamefont
  {Jones}},\ }\bibfield  {title} {\enquote {\bibinfo {title} {Low temperature
  transport in undoped mesoscopic structures},}\ }\href
  {https://doi.org/10.1063/1.3097806} {\bibfield  {journal} {\bibinfo
  {journal} {Applied Physics Letters}\ }\textbf {\bibinfo {volume} {94}},\
  \bibinfo {eid} {172105} (\bibinfo {year} {2009})}\BibitemShut {NoStop}%
\bibitem [{\citenamefont {Pan}\ \emph {et~al.}(2016)\citenamefont {Pan},
  \citenamefont {Baldwin}, \citenamefont {West}, \citenamefont {Pfeiffer},\
  and\ \citenamefont {Tsui}}]{panhig}%
  \BibitemOpen
  \bibfield  {author} {\bibinfo {author} {\bibfnamefont {W.}~\bibnamefont
  {Pan}}, \bibinfo {author} {\bibfnamefont {K.~W.}\ \bibnamefont {Baldwin}},
  \bibinfo {author} {\bibfnamefont {K.~W.}\ \bibnamefont {West}}, \bibinfo
  {author} {\bibfnamefont {L.~N.}\ \bibnamefont {Pfeiffer}},\ and\ \bibinfo
  {author} {\bibfnamefont {D.~C.}\ \bibnamefont {Tsui}},\ }\bibfield  {title}
  {\enquote {\bibinfo {title} {Antilevitation of landau levels in vanishing
  magnetic fields},}\ }\href {https://doi.org/10.1103/PhysRevB.94.161303}
  {\bibfield  {journal} {\bibinfo  {journal} {Phys. Rev. B}\ }\textbf {\bibinfo
  {volume} {94}},\ \bibinfo {pages} {161303} (\bibinfo {year}
  {2016})}\BibitemShut {NoStop}%
\bibitem [{\citenamefont {Pan}\ \emph {et~al.}(2017)\citenamefont {Pan},
  \citenamefont {Kang}, \citenamefont {Baldwin}, \citenamefont {West},
  \citenamefont {Pfeiffer},\ and\ \citenamefont {Tsui}}]{panhigberry}%
  \BibitemOpen
  \bibfield  {author} {\bibinfo {author} {\bibfnamefont {W.}~\bibnamefont
  {Pan}}, \bibinfo {author} {\bibfnamefont {W.}~\bibnamefont {Kang}}, \bibinfo
  {author} {\bibfnamefont {K.~W.}\ \bibnamefont {Baldwin}}, \bibinfo {author}
  {\bibfnamefont {K.~W.}\ \bibnamefont {West}}, \bibinfo {author}
  {\bibfnamefont {L.~N.}\ \bibnamefont {Pfeiffer}},\ and\ \bibinfo {author}
  {\bibfnamefont {D.~C.}\ \bibnamefont {Tsui}},\ }\bibfield  {title} {\enquote
  {\bibinfo {title} {Berry phase and anomalous transport of the composite
  fermions at the half-filled landau level},}\ }\href
  {https://doi.org/10.1038/nphys4231} {\bibfield  {journal} {\bibinfo
  {journal} {Nature Physics}\ }\textbf {\bibinfo {volume} {13}},\ \bibinfo
  {pages} {1168--1172} (\bibinfo {year} {2017})}\BibitemShut {NoStop}%
\bibitem [{\citenamefont {Harris}\ and\ \citenamefont {Lu}(2021)}]{Harris2021}%
  \BibitemOpen
  \bibfield  {author} {\bibinfo {author} {\bibfnamefont {C.~T.}\ \bibnamefont
  {Harris}}\ and\ \bibinfo {author} {\bibfnamefont {T.-M.}\ \bibnamefont
  {Lu}},\ }\bibfield  {title} {\enquote {\bibinfo {title} {A {PtNiGe}
  resistance thermometer for cryogenic applications},}\ }\href@noop {}
  {\bibfield  {journal} {\bibinfo  {journal} {Review of Scientific
  Instruments}\ }\textbf {\bibinfo {volume} {92}},\ \bibinfo {pages} {054904}
  (\bibinfo {year} {2021})}\BibitemShut {NoStop}%
\bibitem [{\citenamefont {Adam}(1969)}]{adam}%
  \BibitemOpen
  \bibfield  {author} {\bibinfo {author} {\bibfnamefont {S.~F.}\ \bibnamefont
  {Adam}},\ }\enquote {\bibinfo {title} {Microwave theory and applicaitons},}\
  \ (\bibinfo  {publisher} {Prentice-Hall},\ \bibinfo {year}
  {1969})\BibitemShut {NoStop}%
\bibitem [{son()}]{sonnet}%
  \BibitemOpen
  \href@noop {} {}\bibinfo {note} {From Sonnet Software, Syracuse, NY;
  www.sonnetsoftware.com.}\BibitemShut {Stop}%
\bibitem [{\citenamefont {Fogler}\ and\ \citenamefont
  {Huse}(2000)}]{foglerhuse}%
  \BibitemOpen
  \bibfield  {author} {\bibinfo {author} {\bibfnamefont {M.~M.}\ \bibnamefont
  {Fogler}}\ and\ \bibinfo {author} {\bibfnamefont {D.~A.}\ \bibnamefont
  {Huse}},\ }\bibfield  {title} {\enquote {\bibinfo {title} {Dynamical response
  of a pinned two-dimensional wigner crystal},}\ }\href
  {https://doi.org/10.1103/PhysRevB.62.7553} {\bibfield  {journal} {\bibinfo
  {journal} {Phys. Rev. B}\ }\textbf {\bibinfo {volume} {62}},\ \bibinfo
  {pages} {7553--7570} (\bibinfo {year} {2000})}\BibitemShut {NoStop}%
\bibitem [{\citenamefont {Eisenstein}, \citenamefont {Pfeiffer},\ and\
  \citenamefont {West}(1994)}]{eisenstein}%
  \BibitemOpen
  \bibfield  {author} {\bibinfo {author} {\bibfnamefont {J.~P.}\ \bibnamefont
  {Eisenstein}}, \bibinfo {author} {\bibfnamefont {L.~N.}\ \bibnamefont
  {Pfeiffer}},\ and\ \bibinfo {author} {\bibfnamefont {K.~W.}\ \bibnamefont
  {West}},\ }\bibfield  {title} {\enquote {\bibinfo {title} {Compressibility of
  the two-dimensional electron gas: Measurements of the zero-field exchange
  energy and fractional quantum hall gap},}\ }\href
  {https://doi.org/10.1103/PhysRevB.50.1760} {\bibfield  {journal} {\bibinfo
  {journal} {Phys. Rev. B}\ }\textbf {\bibinfo {volume} {50}},\ \bibinfo
  {pages} {1760--1778} (\bibinfo {year} {1994})}\BibitemShut {NoStop}%
\bibitem [{\citenamefont {Mehrotra}\ and\ \citenamefont
  {Dahm}(1987)}]{mehrotra}%
  \BibitemOpen
  \bibfield  {author} {\bibinfo {author} {\bibfnamefont {R.}~\bibnamefont
  {Mehrotra}}\ and\ \bibinfo {author} {\bibfnamefont {A.~J.}\ \bibnamefont
  {Dahm}},\ }\bibfield  {title} {\enquote {\bibinfo {title} {Analysis of the
  sommer technique for measurement of the mobility for charges in two
  dimensions},}\ }\href {https://doi.org/10.1007/BF01070654} {\bibfield
  {journal} {\bibinfo  {journal} {Journal of Low Temperature Physics}\ }\textbf
  {\bibinfo {volume} {67}},\ \bibinfo {pages} {115--121} (\bibinfo {year}
  {1987})}\BibitemShut {NoStop}%
\bibitem [{\citenamefont {Burke}\ \emph {et~al.}(2000)\citenamefont {Burke},
  \citenamefont {Spielman}, \citenamefont {Eisenstein}, \citenamefont
  {Pfeiffer},\ and\ \citenamefont {West}}]{burke}%
  \BibitemOpen
  \bibfield  {author} {\bibinfo {author} {\bibfnamefont {P.~J.}\ \bibnamefont
  {Burke}}, \bibinfo {author} {\bibfnamefont {I.~B.}\ \bibnamefont {Spielman}},
  \bibinfo {author} {\bibfnamefont {J.~P.}\ \bibnamefont {Eisenstein}},
  \bibinfo {author} {\bibfnamefont {L.~N.}\ \bibnamefont {Pfeiffer}},\ and\
  \bibinfo {author} {\bibfnamefont {K.~W.}\ \bibnamefont {West}},\ }\bibfield
  {title} {\enquote {\bibinfo {title} {High frequency conductivity of the
  high-mobility two-dimensional electron gas},}\ }\href
  {https://doi.org/10.1063/1.125881} {\bibfield  {journal} {\bibinfo  {journal}
  {Applied Physics Letters}\ }\textbf {\bibinfo {volume} {76}},\ \bibinfo
  {pages} {745--747} (\bibinfo {year} {2000})}\BibitemShut {NoStop}%
\bibitem [{\citenamefont {Ghione}\ and\ \citenamefont
  {Naldi}(1987)}]{ghionenaldi}%
  \BibitemOpen
  \bibfield  {author} {\bibinfo {author} {\bibfnamefont {G.}~\bibnamefont
  {Ghione}}\ and\ \bibinfo {author} {\bibfnamefont {C.}~\bibnamefont {Naldi}},\
  }\bibfield  {title} {\enquote {\bibinfo {title} {Coplanar waveguides for mmic
  applications: Effect of upper shielding, conductor backing, finite-extent
  ground planes, and line-to-line coupling},}\ }\href
  {https://doi.org/10.1109/TMTT.1987.1133637} {\bibfield  {journal} {\bibinfo
  {journal} {IEEE Transactions on Microwave Theory and Techniques}\ }\textbf
  {\bibinfo {volume} {35}},\ \bibinfo {pages} {260--267} (\bibinfo {year}
  {1987})}\BibitemShut {NoStop}%
\bibitem [{onl()}]{onlinecpw}%
  \BibitemOpen
  \href@noop {} {}\bibinfo {note} {See, example,
  https://www.microwaves101.com/calculators/864-coplanar-waveguide-calculator.}\BibitemShut
  {Stop}%
\bibitem [{\citenamefont {Sze}(1981)}]{sze}%
  \BibitemOpen
  \bibfield  {author} {\bibinfo {author} {\bibfnamefont {S.~M.}\ \bibnamefont
  {Sze}},\ }\enquote {\bibinfo {title} {Physics of semiconductor devices},}\ \
  (\bibinfo  {publisher} {Wiley Interscience},\ \bibinfo {year} {1981})\ p.\
  \bibinfo {pages} {849},\ \bibinfo {edition} {2nd}\ ed.\BibitemShut {Stop}%
\bibitem [{\citenamefont {Seeger}(1988)}]{seeger}%
  \BibitemOpen
  \bibfield  {author} {\bibinfo {author} {\bibfnamefont {K.}~\bibnamefont
  {Seeger}},\ }\enquote {\bibinfo {title} {Semiconductor physics: an
  introduction},}\ \ (\bibinfo  {publisher} {Springer},\ \bibinfo {year}
  {1988})\ p.~\bibinfo {pages} {63},\ \bibinfo {edition} {4th}\ ed.\BibitemShut
  {Stop}%
\end{thebibliography}%


%

\end{document}